\documentclass[12pt]{iopart}

\usepackage{epsfig}

\newcommand{\commented}[1]{{}}

\begin{document}

\title[]{Open and Hidden Charm Production at RHIC and LHC\footnote[7]{This 
work was supported in part by the Director, Office of
Energy Research, Division of Nuclear Physics of the Office of High
Energy and Nuclear Physics of the U. S.  Department of Energy under
Contract Number DE-AC03-76SF00098. }}
\author{R. Vogt}

\address{Lawrence Berkeley National Laboratory, Berkeley, CA 94720, USA}

\address{Physics Department, University of California, Davis, CA 95616, USA}

\begin{abstract}

We discuss aspects of open and hidden charm production in hadron-nucleus
collisions at RHIC and LHC energies.  We first discuss the extraction of the
total charm cross section in lower energy collisions and how it compares to
next-to-leading order quantum chromodynamics calculations.  
We then describe calculations of the
transverse momentum distributions and their agreement with the shape of the
measured STAR transverse momentum distributions. We next explain how shadowing
and moderate nuclear absorption can explain the PHENIX $J/\psi$ dAu/$pp$
ratios.

\end{abstract}

\section{Open charm production at RHIC}

Open charm measurements date back to the late 1970s when $D$ 
and $\overline D$ mesons were first detected, completing the
picture of the fourth quark begun when the $J/\psi$ was detected in $p$Be
and $e^+ e^-$ interactions.  The charm quark was postulated to have a mass
between 1.2 and 1.8 GeV, within the regime of perturbative quantum 
chromodynamics (pQCD).  Because of its relatively large mass, it is 
possible to calculate a total $c \overline c$ cross section, not
the case for lighter flavors such as strangeness.  Charm hadrons are usually
detected two ways.  The reconstruction of decays to charged hadrons 
such as $D^0 \rightarrow K^- \pi^+$ (3.8\%) and $D^+ \rightarrow K^- \pi^+ 
\pi^+$ (9.1\%) gives the full momentum of the initial $D$ meson, yielding
the best direct measurement.  Charm can also be detected indirectly via
semi-leptonic decays to leptons
such as $D \rightarrow K l \nu_l$ although
the momentum of the parent $D$ meson remains unknown.  Early measurements
of prompt leptons in beam dump experiments assumed that the density of the dump
was high enough to absorb semi-leptonic decays of non-charm hadrons,
leaving only the charm component.  At modern colliders, it is not possible to
use beam dumps to measure charm from leptons but, at sufficiently
high $p_T$, electrons from charm emerge from hadronic cocktails 
\cite{Kolleggerproc,Hugoproc}.  

Although $D$ mesons are 
usually used to determine the
total $c \overline c$ cross section, other charm hadrons also exist.
The excited $D$ states, $D^*$s, decay primarily to charged and neutral
$D$ mesons.  The charm-strange meson, the $D_s$, decays to charged hadrons
and simultaneously to leptons.  The lowest lying charm baryon
is the $\Lambda_c^+$ which decays primarily to $\Lambda(uds)$ 
but also decays to $pK^- \pi^+$ and to the lepton channel.  The heavier ground
state charm baryons and their excited states ($\Sigma_c$ and higher) decay
through the $\Lambda_c$ channel.  The charm-strange baryons are assumed to be 
a negligible contribution to the total cross section.  A selection of charm
hadrons, their masses, decay lengths and branching ratios to leptons and
charged hadrons are given in Table~\ref{chadtable}.

\begin{table}[htbp]
\begin{center}
\begin{tabular}{ccccc}
$C$ & Mass (GeV) & $c\tau$ ($\mu$m) & $B(C \rightarrow l X)$ (\%)
& $B(C \rightarrow {\rm Hadrons})$ (\%) \\ \hline
$D^+ (c \overline d)$ & 1.869 & 315   & 17.2 & $K^- \pi^+ \pi^+$ (9.1) \\
$D^- (\overline c d)$ & 1.869 & 315   & 17.2 & $K^+ \pi^- \pi^-$ (9.1) \\
$D^0 (c \overline u)$ & 1.864 & 123.4 & 6.87 & $K^- \pi^+$ (3.8) \\
$\overline{D^0} (\overline c u)$ & 1.864 & 123.4 & 6.87 & $K^+ \pi^-$ (3.8) \\
$D^{*\pm}$ & 2.010 &   &   &  $D^0 \pi^\pm$ (67.7), $D^\pm \pi^0$ (30.7) \\
$D^{*0}$   & 2.007 &   &   &  $D^0 \pi^0$ (61.9) \\ \hline
$D_s^+ (c \overline s)$ & 1.969 & 147 & 8 & $K^+ K^- \pi^+$ (4.4), $\pi^+ \pi^+
\pi^-$ (1.0) \\
$D_s^- (\overline c s)$ & 1.969 & 147 & 8 & $K^+ K^- \pi^-$ (4.4), $\pi^+ \pi^-
\pi^-$ (1.0) \\ \hline \hline
$\Lambda_c^+(udc)$   & 2.285 & 59.9 & 4.5 & $\Lambda X$ (35), $pK^- \pi^+$
(2.8) \\
$\Sigma_c^{++}(uuc)$ & 2.452 &      &     & $\Lambda_c^+ \pi^+$ (100) \\
$\Sigma_c^+(udc)$    & 2.451 &      &     & $\Lambda_c^+ \pi^0$ (100) \\
$\Sigma_c^0(ddc)$    & 2.452 &      &     & $\Lambda_c^+ \pi^-$ (100) \\
$\Xi_c^+(usc)$       & 2.466 & 132  &     & $\Sigma^+ K^- \pi^+$ (1.18) \\
$\Xi_c^0(dsc)$       & 2.472 &  29  &     & $\Xi^- \pi^+$ (seen) \\ \hline
\end{tabular}
\end{center}
\caption[]{Ground state charm hadrons with their masses, decay lengths (when
given) and branching ratios to leptons (when applicable) and some prominent
charged hadron decays.}
\label{chadtable}
\end{table}

Extracting the total charm cross section is a non-trivial task.  To go from
a finite number of measured $D$ mesons in a particular decay channel
to the total $c \overline c$ cross section one must: divide by the branching 
ratio; correct for the luminosity, $\sigma_D = N_D/ 
{\cal L}t$; extrapolate to full phase space from the finite detector 
acceptance; divide by two to get the pair cross section from the single $D$s;
and multiply by a correction factor \cite{Mangano} to account for the 
unmeasured charm hadrons.  There are assumptions all along the way.  The most
important is the extrapolation to full phase space.  Before QCD calculations
were available, the data were extrapolated assuming
a power law for the $x_F$ distribution, related to the longitudinal
momentum of the charm hadron by $x_F = p_z/(\sqrt{S}/2) = 2 m_T 
\sinh y/\sqrt{S}$.  The canonical parameterization is $(1 - x_F)^c$ where $c$
was either fit to data over a finite $x_F$ range or simply assumed.  These
parameterizations could lead to large overestimates of the total cross section
when $0<c<2$ was assumed, especially when data were taken only near
$x_F = 0$.  Lepton measurements were more conservative but were typically at
more forward $x_F$.

\subsection{Total $c \overline c$ cross section}

\begin{figure}[htbp] 
\setlength{\epsfxsize=0.95\textwidth}
\setlength{\epsfysize=0.3\textheight}
\centerline{\epsffile{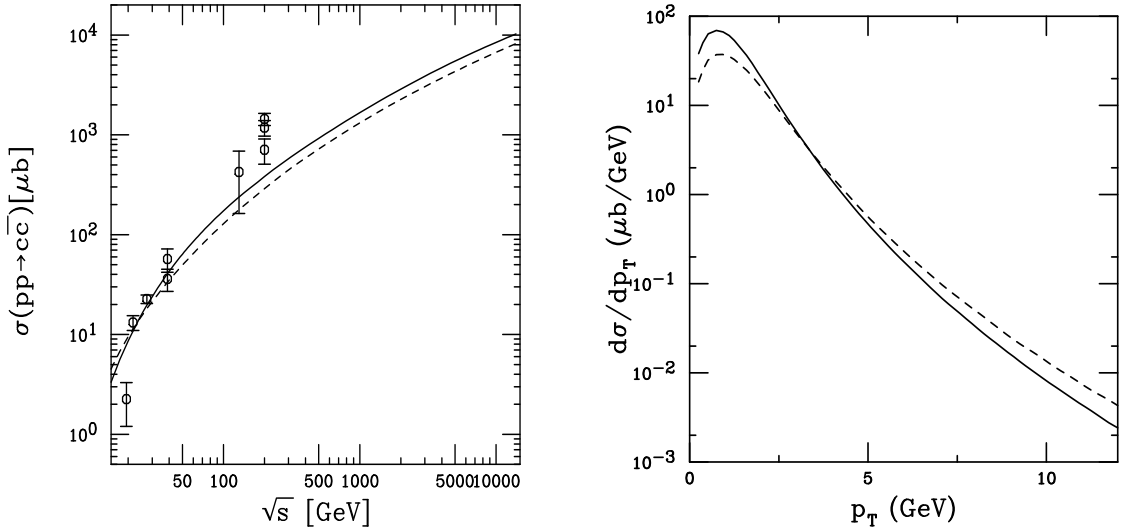}}
\caption[]{The NLO total $c \overline c$ cross sections as a function of
$\sqrt{S}$ (left-hand side)
and charm quark $p_T$ distribution at $\sqrt{S} = 200$ GeV
in the range $|y| \leq 1$ (right-hand side)
in $pp$ interactions.
The curves are MRST HO (solid) with $m = 1.2$ GeV and $\mu^2 =
4m^2$ and GRV98 HO (dashed) with $m = 1.3$ GeV and $\mu^2 = m^2$.
}
\label{totpluspt}
\end{figure}

There has been a great deal of improvement over the last 10-15 years.  
Next-to-leading order (NLO) calculations are used in the phase space 
extrapolation, resulting in considerably less ambiguity in the shape of the
$x_F$ distributions, $d\sigma/dx_F$.  The transverse momentum distributions
are more difficult, as we will discuss later.  To calculate the total
cross section to NLO, scaling functions \cite{NDE} proportional to logs of 
$\mu^2/m^2$ are useful where $\mu$ is the scale of the hard process.  
The hadronic cross section in $pp$ collisions can
be written as
\begin{eqnarray}
\sigma_{pp}(S,m^2) & = & \sum_{i,j = q, \overline q, g} 
\int dx_1 \, dx_2 \, 
f_i^p (x_1,\mu_F^2) \,
f_j^p(x_2,\mu_F^2) \, \widehat{\sigma}_{ij}(s,m^2,\mu_F^2,\mu_R^2)
\label{sigpp}
\end{eqnarray}
where $x_1$ and $x_2$ are the fractional momenta carried by the colliding
partons and $f_i^p$ are the proton parton densities.
The partonic cross sections are
\begin{eqnarray}
\widehat{\sigma}_{ij}(s,m,\mu_F^2,\mu_R^2) & = & 
\frac{\alpha_s^2(\mu_R^2)}{m^2}
\left\{ f^{(0,0)}_{ij}(\rho) \right. \nonumber \\
 & + & \left. 4\pi \alpha_s(\mu_R^2) \left[f^{(1,0)}_{ij}(\rho) + 
f^{(1,1)}_{ij}(\rho)\ln\bigg(\frac{\mu_F^2}{m^2} \bigg) \right] 
+ {\cal O}(\alpha_s^2) \right\}
\,\, .
\label{sigpart}
\end{eqnarray}
with $s$ the squared partonic center of mass energy, $\rho = 4m^2/s$ and 
$f_{ij}^{(k,l)}$ are the scaling functions given to NLO in Ref.~\cite{NDE}.
It is most consistent to assume that the factorization scale, $\mu_F$, and
the renormalization scale, $\mu_R$, are equal, $\mu = \mu_F = \mu_R$.
There is no dependence on the kinematic variables.
Some NNLO calculations are available near threshold, $s = x_1 x_2 S \sim 1.3 \,
(4m^2)$, applicable only for $\sqrt{S} \leq 20-25$ GeV 
\cite{KLMVcc,KVcc}.
The NLO corrections to the leading order (LO) cross sections are relatively
large, $K^{(1)} = \sigma_{\rm NLO}/\sigma_{\rm LO} \sim 2-3$, depending on
$\mu$, $m$ and the parton densities \cite{RVkfac}.  The NNLO corrections are
about as large to next-to-next-to-leading logarithm \cite{KLMVcc} but decrease
to less than $K^{(1)}$ when subleading logs are included \cite{KVcc}.  
This $K$ factor
is large because, in the range $1.2<m<1.8$ GeV, $m <\mu <2m$ with a 5-flavor
QCD scale, $\Lambda_5$, of $0.153$
GeV for the GRV98 HO and $0.22$ GeV for the MRST parton densities,
$0.21 < \alpha_s(c) < 0.4$, nearly a factor of two variation.  
(The larger value corresponds to the
smallest $m$ and $\mu$ values with the larger $\Lambda_5$.)  The situation
improves for bottom where $\alpha_s$ is smaller, $0.16<\alpha_s(b)
< 0.28$, and is quite good for top, $0.092 < \alpha_s(t) < 0.12$.
Instead of presenting a wide range of possible cross sections and emphasizing 
the uncertainties, the approach taken in Ref.~\cite{HPC} has been to 
``fit'' $m$ and $\mu$ for a particular parton density and
extrapolate to higher energies.  The results are compared to some 
of the total cross
section data \cite{Mangano} on the left-hand side of Fig.~\ref{totpluspt}.
The data tend to favor lower values of $m$, $1.2-1.3$ GeV.  The two curves
cross each other because the MRST calculation with $\mu = 2m$ increases faster
at large $\sqrt{S}$ and smaller $x$ due to the stronger QCD evolution of the 
parton densities at the higher scale.  
Although the fixed target results are in good agreement with the
calculations, the PHENIX point \cite{PHENIX130} at 130 GeV, from Au+Au electron
measurements, and the STAR point \cite{STAR}, from a combination of
electron and reconstructed $D$ measurements, are generally above the
calculations.  The STAR point is about
a factor of four over the calculated cross section.  The higher energy
$p \overline p$ data from UA1 \cite{UA1} and CDF \cite{CDF} are in better
agreement with the calculations.  (At these energies, 
the $pp$ and $p \overline p$ cross sections differ by less than 1\% 
for $\sqrt{S} \geq 630$ GeV.)

\subsection{Open charm transverse momentum distributions}

Now we turn to the transverse momentum, $p_T$, distributions.  In this case,
the quark mass is no longer the only scale and $p_T$-dependent logs
also appear.  Thus, to interpolate between
the high $p_T$ scale of $p_T$ and the low $p_T$ scale of $m$, a scale
proportional to $m_T$, the transverse mass, is the natural choice.
The charm quark 
$p_T$ distributions are not strongly dependent on quark mass for $p_T \geq 3$
GeV, as may be expected, where the difference in rate
is $\approx 20$\% between $m = 1.2$ and 1.8 GeV.  The
difference in the total cross sections is almost all at $p_T \leq 3$
GeV.  Changing the scale changes the slope of the $p_T$ distributions.  
The distributions are harder for $\mu = m$ than $\mu = 2m$.
The average $p_T$, $\langle p_T \rangle$, increases with $m$ and is larger for
$\mu = m$.  

More modeling is involved for $D$ mesons in the treatment of 
fragmentation/hadronization and momentum broadening.  If factorization holds
in the final state (universal fragmentation functions) as well as in the
initial state (universal parton distributions) then the fragmentation functions
extracted in $e^+ e^-$ should also be applicable to $pp$ and d$A$.  
However, this assumption does not work well for charm.  The
Peterson function, generally used in hadroproduction codes, reduces the charm
hadron momentum by 30\% relative to the charm quark.  As shown in Huang's talk
\cite{Huangproc}, the Peterson function agrees reasonably well with the $e^+
e^-$ data.  (However, it does not include any scale evolution. In low
$\sqrt{S}$ collisions, the momentum reduction due to fragmentation
can be compensated by intrinsic transverse momentum, $k_T$, 
broadening.  However,
such broadening cannot compensate the $x_F$ distributions,
only marginally affected by $k_T$ smearing.  We have previously shown that the
$D$ meson $x_F$ distributions are consistent with no momentum loss
during charm quark hadronization \cite{VBH2}.)  The exclusive NLO 
$Q \overline Q$ code of Ref.~\cite{MNR} includes
fragmentation and broadening.  This program adds the $k_T$ kick in the
final, rather than the initial state.  The initial
$k_T$ of the partons could have alternatively been given to the entire
final-state system, as is essentially done if applied in the initial state,
instead of to the $Q \overline Q$ pair.  The Gaussian function $g_p(k_T)$,
\begin{eqnarray}
g_p(k_T) = \frac{1}{\pi \langle k_T^2 \rangle_p} \exp(-k_T^2/\langle k_T^2
\rangle_p)
\label{intkt}
\end{eqnarray}
multiplies the parton
distribution functions, assuming the $x$ and $k_T$ dependencies completely
factorize.  If true, it does not matter whether the $k_T$ dependence
appears in the initial or final state.  
There is no difference if the
calculation is LO but at NLO an additional light parton appears in
the final state.  The difference in the two methods is rather small if 
$k_T^2 \leq 2$ GeV$^2$ \cite{Mangano}.  The value $\langle k_T^2 \rangle_p =
1$ GeV$^2$ was used in Ref.~\cite{Mangano}. 

The effects of fragmentation and intrinsic $k_T$ broadening of
$\langle k_T^2 \rangle = 1$ GeV$^2$
compensate each other at $\sqrt{S} = 20$ GeV to give a $D$
meson $p_T$ distribution very similar to that of the charm quark 
\cite{Huangproc}.  However, at RHIC energies, the situation is quite different.
Due to the higher $\langle p_T \rangle$ at larger $\sqrt{S}$, the effect
of broadening is relatively small and cannot compensate for the momentum loss
induced by fragmentation.  Interestingly enough, the STAR $D$ and $D^*$
$p_T$ distribution agrees rather well with the calculated NLO charm
quark distribution, as shown in van Leeuwen's talk \cite{Marcoproc}.  On
the right-hand side of Fig.~\ref{totpluspt}, we show the corresponding $p_T$
distributions at $\sqrt{S} = 200$ GeV for the two sets of parameters in the
total cross section curves on the left-hand side.  The differences in the 
slopes are due to the different scales while the normalization difference is
due to the choice of charm mass and the parton densities --- the MRST densities
generally give a larger cross section due to their larger $\alpha_s$.  However,
the curves need to be scaled up by a factor of four to agree with the STAR
normalization \cite{Marcoproc}, as may be expected from the 
total cross section results.  The shape of the
charm quark $p_T$ distribution at $\sqrt{S} = 1.96$ TeV
also agrees quite well with the CDF data from the
Tevatron \cite{CDF}.  Given the large discrepancy between the pQCD result and
the STAR cross section, it might be surprising that the normalization is also 
in good agreement with the sum of the charged
and neutral $D$ data scaled to include $D_s$ and $\Lambda_c$ production.
No total cross section is available because only charm
hadrons with $p_T > 5$ GeV have been measured so far.  

Other model calculations of charm production at collider energies 
are available.
The FONLL calculation \cite{Cacc} resums logs at $p_T \gg m$,
resulting in a harder charm quark distribution and uses a correspondingly 
harder fragmentation function to get agreement with the CDF data.  It 
underestimates the low $p_T$ STAR data, resulting in an even lower total
cross section than NLO pQCD \cite{ataipriv}.  
A calculation with unintegrated gluon distributions and $k_T$-dependent matrix
elements, assuming saturation behavior at low $x$, has also been made
\cite{Ktuchin}.  However, the $x$ values of the STAR data are not really very
low.  At RHIC, from kinematics $x \sim 0.01$ at $y=1$ and $p_T = 0$, the
highest rapidity measured by STAR by kinematics alone.  In reality, the actual
$\langle x \rangle$ may be higher when weighted by the parton densities.
At higher $p_T$, $x$ is larger still, 
suggesting that the applicability of small $x$ physics for 
charm at RHIC is rather dubious.  

Finally, we would like to discuss reasons why fragmentation does not seem to
factorize for charm, as expected.  Factorization breaking has been suggested
from studies of the $x_F$ distributions of {\it e.g.} $D^+$ and $D^-$ 
production,
particularly in $\pi^- A$ interactions where the $D^-$ is leading relative
to the $D^+$ since the $D^-$ shares a valence quark with the $\pi^-$ while the 
$D^+$ does not.
Several mechanisms such as intrinsic charm \cite{VBH2} and string drag have
been proposed, both of which involve charm quark coalescence with spectators.
Such comoving partons are naturally produced in a hard scattering.  Although
it is not intuitive to expect coalescence to work at high $p_T$, it seems
to do so for charm.

\section{Nuclear dependence of $J/\psi$ production at RHIC and LHC}

We now turn to $J/\psi$ production in d+Au interactions at RHIC.  
Previously, we calculated the effect of shadowing alone on the $J/\psi$
d$A$/$pp$ ratio as a function of rapidity and impact parameter \cite{psidaprl}.
The large $c \overline c$ total cross section also has implications for the
$J/\psi$ yield if $J/\psi$'s arise from $c \overline c$ recombination in
a QGP.  Such a total cross section would suggest significant secondary
$J/\psi$ production at RHIC, leading to enhancement rather than suppression
in central collisions.  There is no evidence for a strong 
regeneration effect in the PHENIX Au+Au data so far, see Thews' talk
\cite{thewsproc}.  

\begin{figure}[htbp] 
\setlength{\epsfxsize=0.5\textwidth}
\setlength{\epsfysize=0.25\textheight}
\centerline{\epsffile{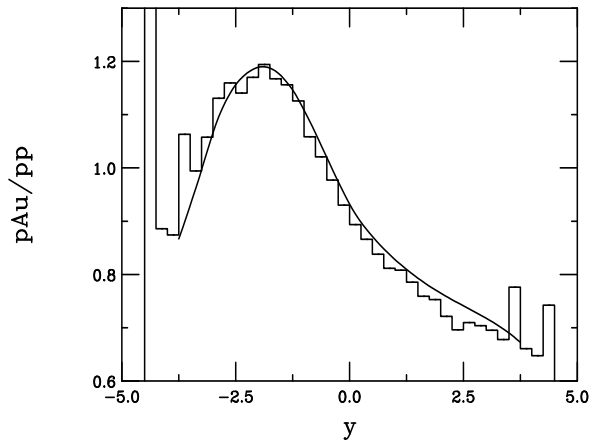}}
\caption[]{ The $J/\psi$ $p$Au/$pp$ ratio at 200 GeV.  We compare the NLO 
(solid histogram, MRST HO) and LO (solid
curve, MRST LO) results using $m = \mu/2 = 1.2$ GeV
with the EKS98 parameterization.  
}
\label{psiratnlo}
\end{figure}

Shadowing, the modification of the parton densities in
the nucleus with respect to the free nucleon, is parameterized as 
$F_i^A(x,\mu^2,\vec b,z) = S^i(A,x,\mu^2,\vec b,z) f_i^p(x,\mu^2)$
in Eq.~(\ref{sigpp}).  We did not discuss the effect of shadowing on
the charm $p_T$ distributions
because the effect at midrapidity is small and, on the logarithmic
scale of the $p_T$ distributions, negligible.  The $J/\psi$
is another story due to the PHENIX muon capability at forward and
backward rapidity.  As shown in Pereira's talk \cite{Hugoproc}, although
the PHENIX $J/\psi$ data are consistent with shadowing alone, the data
are also consistent with nuclear shadowing plus a small absorption
cross section of $1-3$ mb, smaller than that currently obtained in SPS
measurements \cite{Corteseproc}.  We have calculated $J/\psi$ production in
the color evaporation model (CEM) using the same mass and scale as in $c
\overline c$ production but cutting off the invariant mass of the pair at
$4m_D^2$.  The calculations of the d$A$/$pp$
ratios are done at LO to simplify the calculations.  As shown in 
Fig.~\ref{psiratnlo}, the LO and NLO ratios are equivalent.
We have now also implemented nucleon absorption in the calculation,
showing the effect of several absorption and production mechanisms.

\begin{figure}[htbp] 
\setlength{\epsfxsize=0.95\textwidth}
\setlength{\epsfysize=0.3\textheight}
\centerline{\epsffile{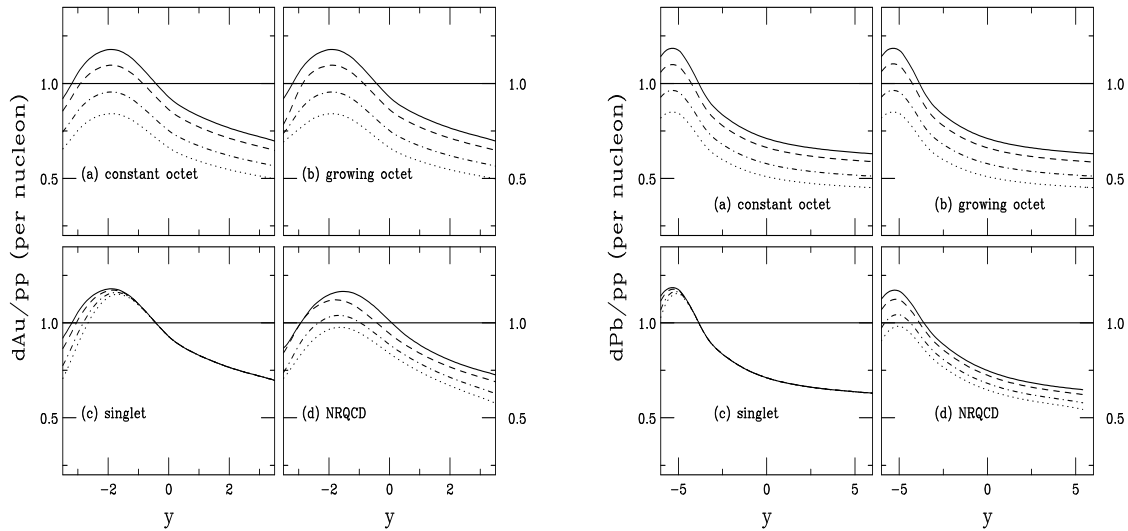}}
\caption[]{The $J/\psi$ d$A$/$pp$ ratio with EKS98 at 200 GeV (left) and
6.2 TeV (right)
as a function of rapidity for (a) constant octet, (b) growing
octet, (c)
singlet, all calculated in the CEM and (d) NRQCD.  For (a)-(c), 
the curves are no 
absorption (solid), $\sigma_{\rm abs} = 1$ (dashed), 3 (dot-dashed)
and 5 mb (dotted).  For (d), we show no absorption 
(solid), 1 mb octet/1 mb singlet 
(dashed), 3 mb octet/3 mb singlet (dot-dashed), and 5 mb octet/3 mb singlet
(dotted).
}
\label{abs}
\end{figure}

To implement nuclear absorption on $J/\psi$ 
production in
$pA$ collisions, the $pN$ production cross section is weighted by the 
survival probability, $S^{\rm abs}$, so that \cite{rvherab} 
\begin{eqnarray} \sigma_{pA} =
\sigma_{pN} \int d^2b \,  \int_{-\infty}^{\infty}\, dz \, \rho_A (b,z) 
S^{\rm abs}(b,z)
\label{sigfull} \end{eqnarray} where $b$ is the impact parameter and $z$ is the
longitudinal production point.  If $S^{\rm abs} = 1$, $\sigma_{pA} = A 
\sigma_{pN}$.  For $S^{\rm abs} \neq 1$, $\sigma_{pA} = A^\alpha \sigma_{pN}$.
We define $S^{\rm abs}$ as 
\begin{eqnarray} S^{\rm abs}(b,z) = \exp \left\{
-\int_z^{\infty} dz^{\prime} \rho_A (b,z^{\prime}) \sigma_{\rm abs}(z^\prime
-z)\right\} \, \, . \label{nsurv} \end{eqnarray} 
The nucleon absorption cross section, $\sigma_{\rm abs}$, depends on where the
state is produced and how far it travels through nuclear matter. 
The effective $A$
dependence is obtained from Eqs.~(\ref{sigfull}) and (\ref{nsurv}) by
integrating over $z'$, $z$, and $b$.  The contribution to the full $A$ 
dependence in $\alpha(x_F)$ from absorption alone is only constant if
$\sigma_{\rm abs}$ is constant and independent of the production mechanism
\cite{rvherab}.
The observed $J/\psi$ yield includes feed down from $\chi_{cJ}$ and $\psi'$
decays, giving
\begin{eqnarray} 
S_{J/\psi}^{\rm abs}(b,z) = 0.58 S_{J/\psi, \, {\rm dir}}^{\rm abs}(b,z) 
+ 0.3 S_{\chi_{cJ}}^{\rm abs}(b,z) + 0.12 S_{\psi'}^{\rm abs}(b,z) 
\, \, . \label{psisurv} 
\end{eqnarray}  
In color singlet production, the final state absorption cross section 
depends on the size of the
$c \overline c$ pair as it traverses the nucleus, allowing absorption to be
effective only while the cross section is growing toward its asymptotic size
inside the target.
On the other hand, if the $c \overline c$ is only produced as a color octet, 
hadronization will occur only after the pair has traversed the target
except at very backward rapidity.  We have considered a constant octet cross
section, as well as one that reverts to a color singlet at backward rapidities.
For singlets, $S_{J/\psi, \, {\rm dir}}^{\rm abs} \neq 
S_{\chi_{cJ}}^{\rm abs} \neq S_{\psi'}^{\rm abs}$ but, with octets,
one assumes that $S_{J/\psi, \, {\rm dir}}^{\rm abs} = S_{\chi_{cJ}}^{\rm abs}
= S_{\psi'}^{\rm abs}$.
As can be seen in Fig.~\ref{abs}, the difference between the constant and
growing octet assumptions
is quite small at large $\sqrt{S}$ with only a small singlet effect 
at $y< -2$ and $-5$ at RHIC and the LHC
respectively.  Singlet absorption is also important only at similar
rapidities and is otherwise not different from shadowing alone.  
Finally, we have also considered a combination of octet and 
singlet absorption in the context of the NRQCD model, see Ref.~\cite{rvherab}
for more details.  The combination of nonperturbative singlet and octet 
parameters changes the shape of the shadowing ratio slightly.  The results are
shown integrated over impact parameter for the EKS98 shadowing parameterization
since it gives good agreement with the trend of the PHENIX data.

We will not discuss the spatial dependence of shadowing and absorption in
any detail here.  The spatial dependence of shadowing alone was discussed
in Ref.~\cite{psidaprl}.  When absorption is included, the trend of the impact
parameter dependence is in agreement with the PHENIX data at $y>0$ (the north
muon arm) but is too weak to describe the strong dependence at $y<0$ (the south
muon arm), see Pereira's talk \cite{Hugoproc}.

\section{Conclusions}

In summary, the RHIC d+Au data on open charm and $J/\psi$ are beginning to
come into their own.  While the QCD calculations agree well with the shape of
the STAR $p_T$ distributions, they underestimate the reported total
cross section.  In contrast, the $J/\psi$ cross section is in relatively
good agreement with QCD predictions \cite{Hugoproc} and the agreement of the
minimum bias data with calculations including shadowing and nucleon absorption 
is quite good.  Work is ongoing to better understand the impact parameter
dependence.

\end{document}